\documentclass[prd,twocolumn,showpacs,preprintnumbers,amsmath,amssymb]{revtex4}

\usepackage{graphicx}
\usepackage{dcolumn}
\usepackage{bm}
\usepackage{mathtools}

\usepackage[breaklinks=true,colorlinks=true,linkcolor=blue,urlcolor=blue,citecolor=blue]{hyperref}

\newcommand{\av}[1]{\langle{#1}\rangle}

\begin{document}

\title[Low- and high-redshift $H_0$ tension resolved]{Emerging spatial curvature can resolve the tension between high-redshift CMB and low-redshift distance ladder measurements of the Hubble constant}

\author{Krzysztof Bolejko}
\affiliation{Sydney Institute for Astronomy, School of Physics, A28, The University of Sydney, NSW, 2006, Australia}%


\begin{abstract}
The measurements of the Hubble constant reveal a tension between high-redshift (CMB) and low-redshift (distance ladder) constraints.
So far neither observational systematics nor new physics has been successfully implemented to explain this tension away. 
This paper present a new solution to the Hubble constant problem. 
The solution is based on the Simsilun simulation (relativistic simulation of the large scale structure of the Universe) with the ray-tracing algorithm implemented.
The initial conditions for the Simsilun simulation were set up as perturbations around the $\Lambda$CDM model. However, unlike in the Standard Cosmological Model (i.e. $\Lambda$CDM model + perturbations), within the Simsilun simulation relativistic and nonlinear evolution of cosmic structures leads to the phenomenon of emerging spatial curvature, where the mean spatial curvature evolves from spatial flatness of the early universe towards slightly curved present-day universe. Consqeuently, the present-day expansion rate is slightly faster compared to the spatially flat $\Lambda$CDM model.
The results of the ray-tracing analysis show that the universe which starts with initial conditions consistent with the Planck constraints should have the Hubble constant 
$H_0 = 72.5 \pm 2.1$ km s$^{-1}$ Mpc$^{-1}$.
When the Simsilun simulation was re-run with no inhomogeneities imposed, the Hubble constant inferred within such a homogeneous simulation
was $H_0 = 68.1 \pm 2.0$ km s$^{-1}$ Mpc$^{-1}$.
Thus, the inclusion of nonlinear relativistic evolution that leads to the emergence of the spatial curvature can explain why the low-redshift measurements 
favour higher values compared to high-redshift constraints  and  alleviate the tension between the CMB and distance ladder measurements of the Hubble constant.
\end{abstract}


\maketitle


\section{Introduction}

For the last 20 years, the $\Lambda$CDM model has been a successful concordance model.
A single, spatially flat model was able to explain properties of high-redshift (early times) and low-redshift (late times) universe. However, with increasing precision of measurements and increasing amount of data,
some tensions between various constraints start to appear \cite{2016IJMPD..2530007B}.
While some of these tensions are subject to observational biases and systematics, which means they are likely to be resolved in the near future, some could pose challenges to the standard $\Lambda$CDM model and point towards various extensions of the standard cosmological model.
One of the most well known example of such tensions is the tension in the measurements of the Hubble constant:
low-redshift measurements of the Hubble constant point towards $H_0 = 73.24 \pm 1.74$ km s$^{-1}$ Mpc$^{-1}$ \cite{2016ApJ...826...56R} whereas the Hubble constant inferred from the CMB (high-redshift) is $H_0 = 67.81 \pm 0.92$ km s$^{-1}$ Mpc$^{-1}$ \cite{2016A&A...594A..13P}.

This paper argues that the $H_0$ tension is a manifestation of rigidity of the FLRW geometry.
Within the FLRW models, if the spatial curvature is flat (the case of the $\Lambda$CDM 
model), it remains flat and does not change with time.
The spatial flatness of the early Universe is predicted by inflation \cite{1981PhRvD..23..347G} and seems to be confirmed by the CMB constraints \cite{2016A&A...594A..13P}. 
Therefore, if our Universe is correctly described (from the early universe till the present day) by the FLRW geometry, then the spatial flatness of the early universe should be preserved.
However, if the geometry of our Universe slightly deviates from the FLRW geometry (for example due to the evolution of cosmic structures \cite{Buchert:2015iva}), then the spatial curvature will not be constrained and spatial flatness may not be preserved \cite{2008CQGra..25s5001B,2011CQGra..28p5004R}.
Direct measurements of the spatial curvature using the low-redshift data (as opposed to fitting the FLRW model to the data) do not place tight constraints on the spatial curvature and allow for large range of possible values, including spatial flatness \cite{2015PhRvL.115j1301R}. 

Understanding the phenomenon of the emerging spatial curvature requires fully relativistic cosmological simulations. However, such simulations are not easy, and so far have not been fully developed \cite{2017IJMPD..2630011B}. 
Cosmological relativistic simulations  based on the Einstein toolkit \cite{2012CQGra..29k5001L}, which implements the BSSN formalism \cite{Bentivegna:2015flc,Mertens:2015ttp,2017PhRvD..95f4028M} have difficulties with shell crossing singularities.
Implementations of post-Newtonian corrections within N-body simulations, do not exhibit problems with shell crossings but face a problem of periodic boundary conditions, which impose a constraint on the global spatial curvature and force it to vanish \cite{2017arXiv170609309A}.
This paper uses a relativistic simulation that is based on the approximation of the `Silent Universes'  --  the Simsilun simulation \cite{simsulun.paper}. The Simsilun simulation starts with perturbations
around the $\Lambda$CDM model. These perturbations are allowed to have a non-zero spatial curvature. Initially, negative curvature of underdense regions is compensated by positive curvature of overdense regions. Once the evolution enters the non-linear regime, the symmetry between overdensities and underdensities is broken, consequently the mean spatial curvature of the universe slowly drifts from zero towards negative curvature induced by cosmic voids. The results of the 
Simsilun simulation indicate that the present-day curvature of our universe is approximately $\Omega_k (z \approx 0) \sim 0.1$, as compared to spatial flatness of the early universe $\Omega_k (z \gg 0) = 0$.

This paper uses the Simsilun simulation (Sec.~\ref{relevo})
and implements the ray-tracing algorithm to generate mock data (Sec.~\ref{raytrac}).
The analysis of the mock catalogues shows that the Hubble constant
inferred from low-redshift data should in fact be higher compared to high-redshift constraints (Sec.~\ref{results}).

\section{Measurements of the Hubble constant}

\subsection{The method of inferring the Hubble constant from CMB }\label{hcmb}

The Hubble constant inferred from the CMB is a highly model-dependent measurement of the present-day expansion rate $H_0$ \cite{2016A&A...594A..13P}.
The parameter $H_0$ is not measured directly but derived from other parameters. 
The standard practice is to fit 6 base parameters of the $\Lambda$CDM model and from them estimate $H_0$.
These six base parameters constitute: 
physical baryon density $\omega_b = \Omega_b h^2$, physical cold dark matter density
$\omega_c = \Omega_c h^2$, optical depth $\tau$,  the amplitude of the dimensional, primordial curvature power spectrum $A_s$, and its spectral index $n_s$.
The last 6th parameter is either the acoustic scale $\theta$ (Planck analysis, \cite{2016A&A...594A..13P}) or the parameter $\Omega_\Lambda$ (WMAP analysis, \cite{2013ApJS..208...19H}).

These two last parameters $\theta$ and  $\Omega_\Lambda$ are not independent from each other. The acoustic scale is defined as a ratio of the sound horizon at decoupling $r_s$ (which depends on physical matter density $\omega_b$, $\omega_c$, and radiation density $\omega_r$) to the angular distance to the last scattering surface $D_A$ (which depends on
physical matter density $\omega_b$, $\omega_c$, radiation density $\omega_r$,  and dark energy density $\omega_\Lambda = \Omega_\Lambda h^2$).
Since, radiation energy density is fixed by the CMB temperature it is not really a free parameter, so apart from $\omega_b$ and $\omega_c$ (which are already listed above) the only free parameter that $\theta$ depends on is $\omega_\Lambda$.

The Hubble constant is then derived:
from the dependence of the shape of the CMB power spectrum on $\Omega_m h^3$,
and the relative height of the acoustic peaks that are sensitive to $\Omega_m h^2$
(Planck analysis, \cite{2016A&A...594A..13P}):
\[ H_0 = 100 \,  {\rm km} \,  {\rm s}^{-1} \,  {\rm Mpc}^{-1} \, \Omega_m h^3 / \Omega_m h^2,\]
or from the condition of the spatial flatness 
(WMAP analysis, \cite{2013ApJS..208...19H}):
\[ H_0 = 100 \,  {\rm km} \,  {\rm s}^{-1} \,  {\rm Mpc}^{-1} \sqrt{\omega_b + \omega_c + \omega_\Lambda}. \]

The physical justification of such a measurement is as follows: the
CMB mostly constrains the physical conditions of the early universe, i.e. 
physical density of baryons, cold dark matter, and radiation.
If one assumes that the evolution of the universe after the decoupling instant is correctly described by the FLRW model then the physical density can be translated to the expansion rate of the present-day universe, $H_0$.
In order to distinguish this parameter from the direct measurement
of the present-day expansion rate (i.e. low-redshift observations of the expansion rate)
let us denote it by $H_0^{ {\rm CMB}}$.
The value of the Hubble constant estimated based on 
measurements obtained by the satellite Planck is $  H_0^{ {\rm CMB}}  = 67.81 \pm 0.92$ km s$^{-1}$ Mpc$^{-1}$ \cite{2016A&A...594A..13P}.

\subsection{The method of inferring the Hubble constant from distance ladder}

At low-redshifts the Taylor expanded FLRW luminosity distance-redshift relation is
\begin{equation}
D_L(z) = \frac{c \,z}{H_0}  \left( 1 + \frac{1}{2} \left[ 1-q_0\right] \, z
- \frac{1}{6} \left[ 1-q_0- 3q_0^2 + j_0\right] \, z^2  \right).
\label{dlzf}
\end{equation}
This low-redshift series is independent of the matter content of the universe,
and the only free parameter apart from $H_0$ are $q_0$ and $j_0$ that can be fixed by the low-redshift data only. In Ref. \cite{2016ApJ...826...56R} these parameters were set to  $q_0 = -0.55$ and $j_0 = 1$ (any realistic variation in $q_0$ and $j_0$ has a minor dependence on the inferred value of $H_0$).

Using the distance modulus to replace the distance with absolute and apparent magnitudes $m$ and $M$
\begin{equation}
m - M = 5 \log_{10} D_L + 25,\label{dismod}.
\end{equation}
the Hubble constant can be written as \cite{2016ApJ...826...56R} 
\begin{equation}
\log_{10} H_0 = \frac{ M + 5\, a + 25}{5}, \label{hdl}
\end{equation}
where 
\begin{eqnarray}
&& a = \log_{10}\left( cz \left\{ 1 + \frac{1}{2} \left[ 1-q_0\right] \, z 
\right. \right. \nonumber \\
&&  \left. \left. 
- \frac{1}{6} \left[ 1-q_0- 3q_0^2 + j_0\right] \, z^2 \right\} \right) - 0.2 \, m. 
\label{aries}
\end{eqnarray}
Thus, to estimate the Hubble constant $H_0$ one needs: redshift $z$, apparent magnitude
$m$, and the absolute magnitude $M$. 
While $z$ and $m$ are directly observable, the absolute magnitude $M$ requires calibrations of standard (or standarisable) candles. 
The calibration can be done using the distance ladder, which uses objects at different distances to calibrate others. In order to distinguish the Hubble parameter derived 
using the distance ladder method (DL) let us denote it
as $H_0^{{\rm DL}}$. The inferred value of the Hubble constant 
based on low-redshfit data is $H_0^{{\rm DL}}= 73.24 \pm 1.74$ km s$^{-1}$ Mpc$^{-1}$ \cite{2016ApJ...826...56R}.

As noted in Sec. \ref{hcmb}, $  H_0^{ {\rm CMB}}  = 67.81 \pm 0.92$ km s$^{-1}$ Mpc$^{-1}$ and so there is a tension between $H_0^{{\rm DL}}$ and $H_0^{ {\rm CMB}}$.
In the next section it will be argued that the main difference between 
$H_0^{{\rm DL}}$ and $H_0^{ {\rm CMB}}$ is not due to observational systematics.
If  the average spatial curvature of our universe evolves from spatial flatness to non-negligible negative values at the present day, then 
one should  in fact expect a difference between $  H_0^{ {\rm CMB}}$ and $H_0^{{\rm DL}}$.

\section{Modelling the relativistic evolution of the Universe}\label{relevo}

\subsection{Silent Universes}\label{silent}

The approximation of the Silent Universes is derived within the 1+3 split \cite{1971grc..conf..104E,2009GReGr..41..581E}. Here, one first introduces the comoving gauge with the velocity field $u^a \sim \delta^a_{\,0}$, and assumes that the gravitational field is sourced by irrotational and insulated dust. Then applying the energy-momentum conservation $T^{ab}{}_{;b} =0$, the Ricci identities $u_{a;d;c}-u_{a;c;d} = R_{abcd} u^b$, and the Bianchi identities $R_{ab[cd;e]} = 0$, and finally assuming  with the magnetic part of the Weyl tensor vanish, one reduces the Einstein equations
to only 4 equations, which describe the evolution of dust (with matter density $\rho$), its velocity filed (with expansion rate $\Theta$ and shear $\Sigma$) and the Weyl curvature ${\cal W}$ \cite{1995ApJ...445..958B,1997CQGra..14.1151V}

\begin{eqnarray}
&& \dot \rho = -\rho\,\Theta, \label{rhot}\\
&& \dot \Theta = -\frac{1}{3}\Theta^2-\frac{1}{2}\,\kappa \rho-6\,\Sigma^2 + \Lambda,\label{thtt}\\
&& \dot \Sigma = -\frac{2}{3}\Theta\,\Sigma+\Sigma^2-{\cal W},\label{shrt}\\
&& \dot{ {\cal W}} = -\Theta\, {\cal W} -\frac{1}{2}\,\kappa \rho\,\Sigma-3\Sigma\,{\cal W},\label{weyt} 
\end{eqnarray}
where $\kappa = 8 \pi G/c^4$. In addition to these equations, the evolution of the volume $V$ of the fluid's element is given by
\begin{eqnarray}
\dot{V} = V \Theta.\label{volt}
\end{eqnarray}
Apart from the evolution equations there are also spatial constraints. However, if these constraints are initially satisfied, they will be preserved in the course of the evolution \cite{1997CQGra..14.1151V,1997PhRvD..55.5219M}. Thus, once the initial conditions are properly set up, the evolution of a relativistic system can be evaluated based on the above equations only.
Finally, there is also the ``Hamiltonian'' constraint which can be used to evaluate the spatial curvature
\begin{equation}
\frac{1}{6} {\cal R}= \frac{1}{3}\,\kappa   \rho + \Sigma^2  - \frac{1}{9} \Theta^2 + \frac{1}{3} \Lambda. \label{hamcon}
\end{equation}

\subsubsection{FLRW limit}\label{FLRWlimit}

In the limit of spatial homogeneity and isotropy the Silent Universes reduces to the FLRW models. 
The condition of spatial homogeneity and isotropy implies that shear vanishes at every point in space and time, hence $\Sigma \equiv 0 \equiv \dot{\Sigma}$, which also implies ${\cal W} =  0$ and $\dot{{\cal W}} =  0$. Thus, the last two equations of the Silent Universe are trivial. The condition of spatial homogeneity and isotropy also leads to $\Theta \to 3 \dot{a} / a$, where the function $a(t)$ depends on time only and is the FLRW scale factor (thus the scalar of the expansion $\Theta$ is 3 times the Hubble parameter). Then the first equation of the Silent Universe reduces to $\rho = \rho_i a^{-3}$, and the second 
reduces to 

\begin{equation}
3 \frac{\ddot{a}}{a} = - \frac{1}{2} \kappa \rho  +  \Lambda,  \label{fes1}
\end{equation}
which is the first Friedmann equation. In the limit of spatial homogeneity and isotropy,
the spatial curvature reduces to ${\cal R} \to 6 k/a^2$ and the Hamiltonian constraint becomes
\begin{equation}
  3 \frac{\dot{a}^2}{a^2} = \kappa  \rho  - 3 \frac{k}{a^2}   + \Lambda  \label{fes2},
\end{equation}
which is  the first Friedmann equation.

\subsection{Simsilun simulation}\label{simsilun}

The evolutionary equations of the  Silent Universe, i.e. (\ref{rhot})--(\ref{weyt}) 
has been implemented in the code \textsl{simsilun} \footnote{\url{https://bitbucket.org/bolejko/simsilun}}.  
The description of the code, equations, and applications are described in the `Methods Paper' \cite{simsulun.paper}.
The Methods Paper describes how one can use the 
 Millennium simulation \cite{2005Natur.435..629S,2009MNRAS.398.1150B,2013MNRAS.428.1351G}
to set up the initial condition for the code \textsl{simsilun}.
The initial conditions are set up using the smoothed density field 
of the  Millennium simulation stored in the  MField database \footnote{The MField database is accessible via the German Astrophysical Virtual Observatory \url{
http://gavo.mpa-garching.mpg.de/MyMillennium} with the SQL query \texttt{select * from MField..MField}.}.

\begin{figure}
\begin{center}
\includegraphics[scale=0.68]{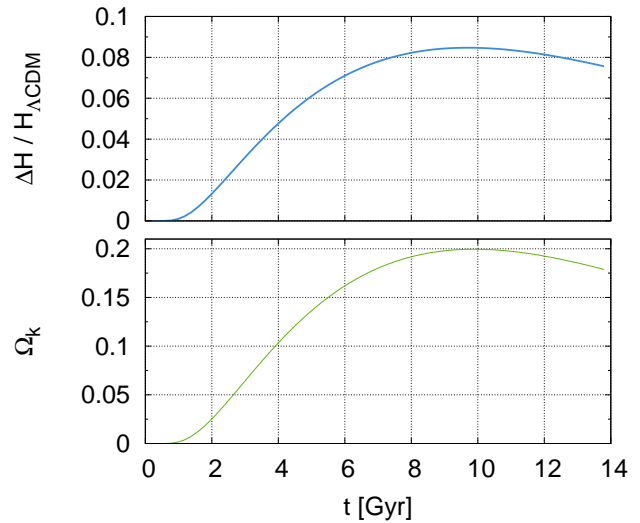}
\end{center}
\caption{Evolution of the global (mean) expansion rate (Upper panel) and the spatial curvature (Lower panel) within the Simsilun simulation. The initial conditions for the  Simsilun simulation has been setup using density fluctuations from the Millennium simulation imposed on the Planck's $\Lambda$CDM model at $z_i=80$. As long as perturbations remain within the linear regime ($t<1$ Gyr) the mean evolution follows the $\Lambda$CDM model. Once the system enters non-linear regime the spatial curvature emerges
and the expansion rate slightly increases compared to the $\Lambda$CDM model.}
\label{fig1}
\end{figure}

In this paper we apply a slight modification of the \textsl{Simsilun simulation}
discussed in the Methods Paper \cite{simsulun.paper}.
The MField consists of $256^3$ cells, that contain information about the matter density field
smoothed with Gaussian kernel of radius $1.25\,h^{-1}$ Mpc, $2.5\,h^{-1}$ Mpc,
$5\,h^{-1}$ Mpc, and $10\,h^{-1}$ Mpc. 
Unlike in the Methods Paper, where the smoothing scale was $2.5\,h^{-1}$ Mpc, here we use the matter field smoothed with $1.25\,h^{-1}$ Mpc radius, as it reproduces the parameter $\sigma_8$ more accurately --- the smoothing decreases the variance of the density filed so the larger the smoothing radius the smaller the parameter $\sigma_8$: with $1.25\,h^{-1}$ Mpc smoothing kernel the parameter $\sigma_8$ is underestimated by less than $2\%$, with
$2.5\,h^{-1}$ Mpc by $13\%$, $5\,h^{-1}$ Mpc by $30\%$, and with $10\,h^{-1}$ Mpc by almost $60\%$.
The second change, compared to the Methods Paper is the change of the
background cosmology. The Simsilun simulation described in the Methods Paper is 
based on the WMAP1 cosmology, just as the  Millennium simulation.
Here we assume that the background model is the  Planck's $\Lambda$CMD model, and we use it to set up the initial conditions for the Simsilun simulation. The initial background density $\bar{\rho}_i$ and the initial background's expansion rate $\bar{\Theta}_i$ are 

\begin{eqnarray}
&& \bar{\rho}_i = \Omega_m \frac{3 H_0^2}{8 \pi G} (1+z_i)^3 = 
\omega_m \frac{3}{8 \pi G} (1+z_i)^3 \, H_{100}^2 \\
&& \bar{\Theta}_i = 3 \, H_{100}\sqrt{ \omega_m (1+z_i)^3 + \omega_\Lambda },
\end{eqnarray}
where $H_{100} = 100 \, {\rm km} \, {\rm s}^{-1}\, {\rm Mpc}^{-1}$, $\omega_m = \Omega_m h^2= 0.1415$, and $\omega_\Lambda = \Omega_\Lambda h^2 = 0.3182$  \cite{2016A&A...594A..13P}.
We then use the 
 initial perturbations at $z_i = 80$, which follow from the Millennium's snapshot no 1,
and superimpose them onto the Planck's $\Lambda$CMD background model ($\bar{\rho}_i$ and 
$\bar{\Theta}_i$).
This serves as the initial conditions for our new simulation, that is based on evolving 16,777,216 worldlines (i.e. $256^3$ cells) using eqs. (\ref{rhot})--(\ref{volt}) up to $z = 0$.

As discussed in the Methods Paper, the result of the evolution of the Silent Universe
is emergence of the spatial curvature. The emergence of the spatial curvature is associated with the increase of the mean expansion rate, which is presented in Fig. \ref{fig1}.
The mean expansion rate is defined as the volume average 

\begin{equation}
H_{\cal D} =  \frac{1}{3} \av{ \Theta }_{\cal D} = 
\frac{1}{3} 
 \frac{ \sum_n \, \Theta_n \, V_n} {\sum_n V_n}, \label{volav}
\end{equation}
where the domain ${\cal D}$ is the whole domain of the Simsilun simulation,
$\Theta_n$ is the expansion rate of a single worldline/cell, and $V_n$ is its volume,
and  ${\sum_n V_n}$ is  volume of the entire 
domain  of the Simsilun simulation. The parameter
$\Omega_k^{\cal D}$ of the spatial curvature is
\begin{equation}
\Omega_k^{\cal D} = -\frac{ \av{\cal{R}}_{\cal D} } { 6 H_{\cal D}^2 }.
\end{equation}
In the limit of spatial homogeneity and isotropy, each cell has the same expansion rate $\Theta_i$ and thus the same volume (cf. (\ref{volt})). Consequently, $H_{\cal D} \to H_{\Lambda CDM}$ and $\Omega_k^{\cal D}  \to - k / \dot{a}^2$. Therefore if $k=0$ then 
also $\Omega_k^{\cal D} = 0$. Within the regime of linear perturbations,
all quantities can be expressed in terms of density perturbations $\Delta \rho$. Since the average of linear perturbations vanishes, thus the average expansion coincides with the background expansion rate, i.e.  $H_{\cal D} = H_{\Lambda CDM}$.
Similarly, the average spatial curvature, within the linear regime is flat  $\Omega_k^{\cal D} = 0$.
It is only in the nonlinear regime when the
spatial curvature emerges and the expansion rate increases compared to the 
$\Lambda$CDM model. 
This is presented in Fig. \ref{fig1} where in the nonlinear regime ($t> 1$ Gyr)
both spatial curvature and the expansion rate deviate from the $\Lambda$CDM model.
It is interesting to note that in the dark energy dominated epoch ($t> 10$ Gyr)
both the spatial curvature and expansion rate do asymptotically approach the $\Lambda$CDM model; this phenomenon is know as the ``cosmic no-hair'' conjecture 
\cite{2017arXiv171204041B}.

However, it needs to be stressed  that the expansion rate $H_{\cal D}$ (presented in Fig.~\ref{fig1}) is not the same as the Hubble constant inferred from the distance ladder $H_0^{{\rm DL}}$ --- it is only in the FLRW limit where the
expansion rate and the slope of the distance-redshift relation are equivalent to each other \cite{2010GReGr..42.2453K}. Therefore, in order to estimate the Hubble parameter based on 
the distance ladder $H_0^{{\rm DL}}$  one needs to implement the ray-tracing
method to the Simsilun simulation, which is described in the next section.

\section{Light propagation}\label{raytrac}

\subsection{Distance and redshift}

Apart from the evolution of the universe the ray-tracing is implemented within the 
 Simsilun simulation. The light propagation is based on the Sachs optical equations \cite{1961RSPSA.264..309S}. The angular diameter distance $D_A$ follows from

\begin{equation}
\frac{{\rm d^2} D_A}{{\rm d} s^2} = -  \left( \sigma^2 + \frac{1}{2} R_{a b} k^a k^b \right)  D_A, \label{dsr}
\end{equation}
where $\sigma$ is the shear of the null bundle $k^a$, and  $s$ is the affine parameter. The redshift follows from

\begin{equation}
 \frac{ {\rm d} z } {  {\rm d} s } = 
\left(  \frac{1}{3} \Theta + \Sigma_{ab} \, n^a n^b \right) (1+z)^2,
\end{equation}
where $\Sigma_{ab}$ the shear of the matter filed and $n^a$ is a unit vector in the direction of propagation.
For comoving dust $R_{a b} k^a k^b = \rho (1+z)^2$ and for non-extreme cases (strong lensing) the null shear does not affect the distance-redshift relation  \cite{2012JCAP...05..003B}. Additionally, since there is no prefer direction,
the average contribution from matter shear to the to distance relation vanishes,
as it only contributes via the trace \cite{2010JCAP...03..018R}.

Solving the above equations within the Simsilun simulation we find the relation between the angular diameter distance and redshift. Then using the reciprocity theorem \cite{1971grc..conf..104E} the luminosity distance is 
\begin{equation}
D_L = (1+z)^2 \, D_A.
\end{equation}

\subsection{Generating the mock catalogues}\label{mocks}

In Ref. \cite{2016ApJ...826...56R} the Hubble constant estimated based on low-redshfit data was inferred using a two-stage analysis. First, the parameter $a$ 
of eq. (\ref{aries})
was inferred from 217 supernova Ia with redshifts $0.0233 < z < 0.15$. Then various anchors were used to perform a
simultaneous fit of supernova and Cepheid data to infer $M$, which 
in turn via  (\ref{hdl}) constrained the Hubble constant $H_0^{{\rm DL}}$.

Within the Simsilun simulation the implemented ray-tracing algorithm  
provides the distances, consequently the last step of the calibration of $M$ for the Simsilun simulation is not needed and the Hubble parameter $H_0^{{\rm DL}}$ can be estimated from
\begin{eqnarray}
&& \log_{10} H_0^{{\rm DL}} = \log_{10}\left( cz \left\{ 1 + \frac{1}{2} \left[ 1-q_0\right] \, z 
\right. \right. \nonumber \\
&&  \left. \left. 
- \frac{1}{6} \left[ 1-q_0- 3q_0^2 + j_0\right] \, z^2 \right\} \right) - \log_{10}  D_L,
\label{Hsig}
\end{eqnarray}
or in terms of the distance modulus from
\begin{eqnarray}
&& \log_{10} H_0^{{\rm DL}} = \log_{10}\left( cz \left\{ 1 + \frac{1}{2} \left[ 1-q_0\right] \, z 
\right. \right. \nonumber \\
&&  \left. \left. 
- \frac{1}{6} \left[ 1-q_0- 3q_0^2 + j_0\right] \, z^2 \right\} \right) - 0.2 \, \mu + 5.
\label{Hsigmu}
\end{eqnarray}
where the distance modulus is $\mu = 5 \log_{10} D_L + 25$.

\begin{figure}
\begin{center}
\includegraphics[scale=0.68]{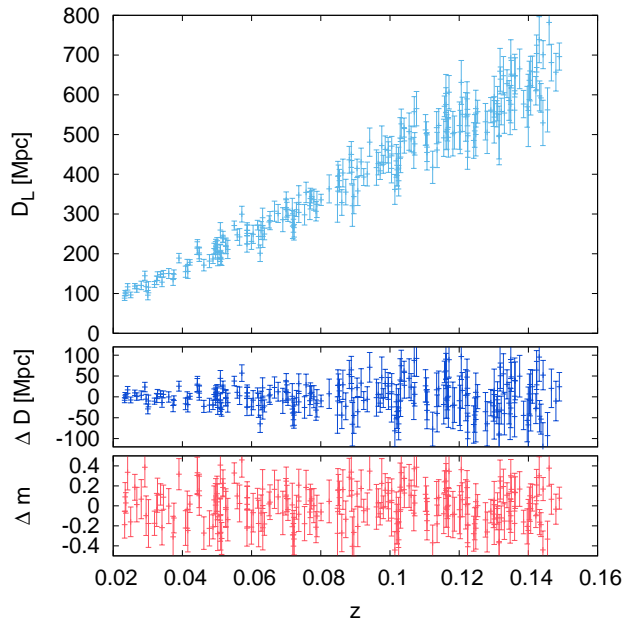}
\end{center}
\caption{A single mock supernova catalogue generated within the Simsilun simulation. 217 generated distance-redshift relations has been Gaussian scattered using uncertainties from the Union2.1 data set. Upper panel shows the luminosity distance $D_L$; Middle panel shows distance residuals  $\Delta D_L$ from the  best-fit distance-redshift relation (\ref{dlzf}); Lower panel shows residuals in brightness $\Delta \mu = 5 \log_{10} (1+\Delta D_L/D_L)$.}
\label{fig2}
\end{figure}

To estimate the Hubble constant $H_0^{{\rm DL}}$ within the Simsilun simulation we generate 217 light rays with redshift $0.0233 < z_{0,i} < 0.15$, and calculate the luminosity distance $D_{0,i}$. 
To apply some realistic uncertainties we use the Union2.1 set \cite{2012ApJ...746...85S}.
We take uncertainties and covariance matrix from the Union2.1 dataset\footnote{The supernova data as well as the covariance matrix for the Union2.1 set is accessible via the website of the Supernova Cosmology Project \url{http://supernova.lbl.gov/Union/}.}
 for 217 supernova with $z<0.2$ (Union2.1 consists of 580 supernova with redshift up to $z = 1.414$) and apply them to the Simsilun simulation distances. While this procedure is not ideal, it does provide `realistic' uncertainties, that can be applied to the `ideal' data generated using the  Simsilun simulation.
First, the uncertainty in the distance modulus is transform to uncertainty in each distance $D_{0,i}$ 
\[ \Delta D_i = 0.2 \Delta \mu_i \, D_{0,i} \, \log_{10} 10, \]
and then the distances are Gaussian scattered 
\[ D_{L,i} = {\cal N} (\mu = D_{0,i}, \sigma=\Delta D_i), \]
where ${\cal N} (\mu = D_{0,i}, \sigma=\Delta D_i)$
is a random number drawn from a Gaussian distribution
whose mean value is $D_{0,i}$ and standard deviation equal to distance uncertainty $\Delta D_i$.
Once the mock catalogue is generated, we perform the MCMC analysis. The likelihood at each step is evaluated based on eq. (\ref{Hsigmu}) with the covariance matrix taken from the Union2.1 set. The MCMC analysis allows to estimate the mean, as well as, uncertainties in 
$H_0^{{\rm DL}}$  while treating the parameters $q_0$ and $j_0$ as the nuisance parameters.
An example of a single mock catalogue (next sections considers multiple mocks)
together with the residuals from the best-fit are presented in Fig. \ref{fig2}.

\section{Results}\label{results}

\begin{figure}
\begin{center}
\includegraphics[scale=0.68]{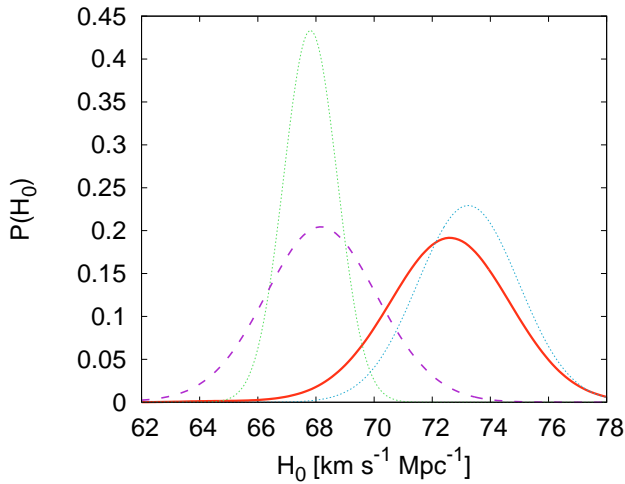}
\end{center}
\caption{The Hubble constant evaluated within the Simsilun simulation (red solid line). 
The constraints are inferred from the slop of the distance-redshift relation 
(\ref{Hsigmu}) and result with $H_0^{{\rm DL}} =  72.5 \pm 2.1$ km s$^{-1}$ Mpc$^{-1}$.
 If the Simsilun simulation is run with no inhomogeneities imposed (i.e. the FLRW case) then the Hubble constant inferred from the slop of the distance-redshift is $H_0^{{\rm DL}} = 68.1 \pm 2.0$ km s$^{-1}$ Mpc$^{-1}$ (purple dashed line). For comparison, the green dotted line (on the left) presents the Gaussian profile with the mean $67.81$ and the standard deviation  $0.92$ (cf. CMB constraints \cite{2016A&A...594A..13P}) and the blue dotted line (on the right)
shows a Gaussian profile with the mean $73.24$ km s$^{-1}$ Mpc$^{-1}$ and the standard deviation  $1.74$ km s$^{-1}$ Mpc$^{-1}$ (cf. distance ladder constraints \cite{2016ApJ...826...56R}).}
\label{fig3}
\end{figure}

The results for the Hubble constant $H_0^{{\rm DL}}$ estimated using
eq. (\ref{Hsigmu}) based on the ray-tracing within the Simsilun simulation
are presented in Fig. \ref{fig3}. The results include the cosmic variance 
which was estimated using $10,000$ mock catalogues with random observers.
The Hubble constant is $H_0^{{\rm DL}} = 72.5 \pm 2.1$ km s$^{-1}$ Mpc$^{-1}$ and its pdf is presented with a red solid line in Fig. \ref{fig3}. It should be noted that the initial conditions for the 
Simsilun simulation were set up using the Planck data. When the Simsilun simulation
was re-run with no inhomogeneities imposed, the Hubble constant inferred within such a 
homogeneous simulation\footnote{A homogeneous 
Simsilun simulation is the one where the initial density contrast vanish everywhere $\delta_i = 0$, this means that $\Sigma = 0$ and ${\cal W} = 0$ and so the system is equivalent to the FLRW system (cf.  Sec. \ref{FLRWlimit}), however the evolution is still traced by solving eqs. (\ref{rhot})--(\ref{volt}).} (using the mock catalogues 
generated as described in Sec. \ref{mocks}) was found to be  
$H_0^{{\rm DL}} = 68.1 \pm 2.0$ km s$^{-1}$ Mpc$^{-1}$. The pdf of
the Hubble constant inferred from a homogeneous Simsilun simulation 
is presented with a purple dashed line in 
Fig. \ref{fig3}. This shows that relativistic non-linear evolution 
of a cosmic system, which allows for the emergence of the spatial curvature can solve the problem of the tension between 
high-redshift (CMB) and low-redshift (distance ladder) measurements of $H_0$. 
For comparison
the pdfs of these two measurements are presented in Fig. \ref{fig3} using dotted lines \footnote{To be precision, the dotted lines shows 
Gaussian profiles: one with the mean $\mu = 67.81$ and standard deviation $\sigma = 0.92$ (CMB), and the other one with $\mu = 73.24$ and $\sigma = 1.74$ (distance ladder).}.

\section{Discussion on the origin of the effect}

The reason why the Simsilun simulation predicts a higher expansion rate
has been partially explained in Sec. \ref{simsilun}.
However, there is an additional subtlety that needs to be discussed in regards
to the results obtained in Sec. \ref{results}.

The expansion rate of space follows from  eq. (\ref{thtt}).
As seen from eq. (\ref{thtt}), in the linear regime the only factor
that causes the departure from the background's expansion rate $\bar{\Theta}$ is the 
fluctuation in the density field $\rho = {\bar \rho} + \Delta \rho$.
Since the global average of density fluctuations vanishes, i.e. $\int {\rm d} V \, \Delta \rho = 0$, a faster expansion rate of voids (where $\Delta \rho < 0$) is compensated by a slower expansion rate of overdense regions (where $\Delta \rho > 0$), consequently the average expansion rate coincides with the background's expansion rate.

Once the evolution becomes nonlinear, the symmetry between underdense and overdense regions starts to break. On the one side, as the underdense regions become emptier they expand faster; on the other side, the build-up of the shear $\Sigma^2$ within overdense regions slows down their expansion rate more efficiently than just the density perturbations alone. 
Consequently, the average expansion rate is faster compared to the $\Lambda$CDM model, i.e. 
$H_{\cal D} > (\dot{a}/a)_{\Lambda CDM}$.

For comparison, this effect is not present within the standard N-body simulations.
Within the N-body simulations, matter is inhomogeneously distributed and
even though one could map these fluctuations onto the shear and Weyl curvature,
these quantities do not affect the overall expansion rate of the universe.
Within the standard N-body simulations the expansion rate is
given by  eq. (\ref{fes2}) and it is uniform everywhere,
i.e. $H_{0} = (\dot{a}/a)_{\Lambda CDM}$.

Within the Simsilun simulation 
cosmic voids occupy more volume than other regions,
as a result if one picks a random line of sight, then along such a line of sight light most likely propagates through underdense regions.
Although it sounds similar, this is not the Dyer-Roeder  effect \cite{1972ApJ...174L.115D,1973ApJ...180L..31D}, which is related to pure density fluctuations \cite{2011MNRAS.412.1937B}. Here the effect is related to propagation through regions that expand faster than the background \cite{2012MNRAS.426.1121C,2012MNRAS.419.1937M}.
Most importantly though, the effect reported in Sec. \ref{results}
should not be mistaken and contributed to `insufficient randomisation' of the line of sights
\cite{2008PhRvD..78h3511V,2009PhRvD..80l3020K}.
On the contrary, had the insufficient randomisation been the issue,
i.e. had we chosen light rays that propagate only through underdense regions then 
$H_0^{{\rm DL}}$
would have been up to 30\% higher (instead of  6.5\% higher) compared to the 
$\Lambda$CDM model.
The fact that the Hubble constant
 inferred from the distance-redshift relation $H_0^{{\rm DL}}$
is so similar to the average expansion rate of space $H_{\cal D}$ 
empirically confirms results obtained in Refs.
\cite{2009JCAP...02..011R,2012JCAP...05..003B,2016MNRAS.455.4518K},
which suggest that the average distance-redshift relation should
follow the average expansion rate.

As seen from Fig. \ref{fig1}, once the evolution becomes nonlinear, the average expansion rate starts to deviate from the background's $\Lambda$CDM model, thus
the present-day expansion rate inferred from the distance-redshift relation 
should be higher than the Hubble constant inferred from the conditions of the early universe, i.e. $H_0^{{\rm DL}} > H_0^{{\rm CMB}}$. This expectation is indeed confirmed by the results presented in Fig. \ref{fig3}.

\section{Conclusions}\label{conclusions}

The history of measurements of the Hubble constant
shows how its value, at various stages of time, 
was susceptible  to number of observational biases.
During the 20th century the value of the Hubble constant was subject to 
number of changes: misclassification of Cepheids, confusion between stars and HII regions, and especially the Malmquist bias \cite{2006RvMA...19....1T} often led to overestimation of its value. However, when at the turn of the century the HST Key Project settled its value to $72 \pm 8$ km s$^{-1}$ Mpc$^{-1}$ \cite{2001ApJ...553...47F} it seemed that most systematics got under control. Yet when 10 years later, the 7-year WMAP data pointed towards $H_0^{{\rm CMB}} = 70.2 \pm 1.4$  km s$^{-1}$ Mpc$^{-1}$ \cite{2011ApJS..192...18K} and the distance ladder method towards $H_0^{{\rm DL}} = 73.8 \pm 2.4$ km s$^{-1}$ Mpc$^{-1}$ \cite{2011ApJ...730..119R}, the debate on the Hubble constant got revived. The tension between high-redshift  measurements (CMB)
and low-redshift (distance ladder) got further widen with the Planck measurements, which  constrained it to $H_0^{{\rm CMB}} = 67.3 \pm 1.2$  km s$^{-1}$ Mpc$^{-1}$ \cite{2014A&A...571A..16P}. 
The inconsistency between these measurements seems to be statistically significant and does not seem to appear simply because we have two different types of measurements 
\cite{2017PhRvD..96h3532L}. This suggest there must be some mechanism behind this inconsistency, be it either unaccounted systematics or some physical phenomenon.

The issue of systematics in the distance ladder method resurfaced when it was pointed out that various assumptions regarding the calibrations can shift the value of the Hubble constant
by 2 km s$^{-1}$ Mpc$^{-1}$ \cite{2014MNRAS.440.1138E}. Given the lack of compelling evidence for `new physics' \cite{2014ApJ...794..135B} it seemed likely that once again (as often in the past) the systematics were to be blamed for overestimating the Hubble constant. However, a careful analysis of the distance ladder using multiple anchors resulted with $H_0^{{\rm DL}} = 73.24 \pm 1.74$ km s$^{-1}$ Mpc$^{-1}$ \cite{2016ApJ...826...56R}, which confirmed the tension. The tension was further solidified 
with the latest Planck measurements, which set the high-redshift Hubble constant to
$H_0^{{\rm CMB}}= 67.81 \pm 0.92$ km s$^{-1}$ Mpc$^{-1}$ \cite{2016A&A...594A..13P}.

This paper explored an extension of the $\Lambda$CDM model. This extension does not require any new physics in terms of the dark sector (e.g. evolving dark energy or interacting dark matter) or in terms of modification of gravity.
The solution that this paper provides is more prosaic. It relies on the fact that the Einstein equations are nonlinear and therefore the evolution
of an inhomogeneous nonlinear system that only in a statistical sense is homogeneous and isotropic (i.e. after averaging over sufficiently large domains) is not exactly the same as the evolution of an exactly homogeneous and isotropic system \cite{Buchert:2015iva}.

This paper uses the Simsilun simulation that solves the Einstein equations within the approximation of the Silent Universes \cite{simsulun.paper}.
Within the framework of the Simsilun simulation  the
spatial curvature evolves from spatial flatness of the early universe
to a slightly negative values at the present day. A slight increase of the spatial curvature speeds up the expansion rate, which is presented in Fig.~\ref{fig1}. 
The implementation of the ray-tracing algorithm within the Simsilun simulation  allowed to generate mock supernova data, which were used to estimate the low-redshift Hubble constant $H_0^{{\rm DL}}$ directly from the distance-redshift relation. The initial conditions for the Simsilun simulation has been setup in the early universe around the Planck's $\Lambda$CDM model \cite{2016A&A...594A..13P}. 10,000 mock catalogues has been generated (an example of a mock catalogue with uncertainties imposed from Union2.1 set \cite{2012ApJ...746...85S} is presented in Fig.~\ref{fig2}). The analysis showed that the phenomenon of 
emergence of the spatial curvature can solve the Hubble constant problem. 
As shown in Fig. \ref{fig3}, if the evolution of the universe follows exactly the equation of a purely homogeneous and isotropic universe (FLRW case) 
then the tension between the low-redshift and high-redshift Hubble constant appears. If however, relativistic corrections due to nonlinear cosmic evolution are included, then the tension is alleviated.
The results of the evolution and ray-tracing algorithms within the Simsilun simulation
show that staring from the initial conditions as prescribed by the Planck satellite
($\omega_m = \Omega_m h^2= 0.1415$, and $\omega_\Lambda = \Omega_\Lambda h^2 = 0.3182$  \cite{2016A&A...594A..13P}) the present-day expansion rate should in fact be 
$H_0 = 72.5 \pm 2.1$ km s$^{-1}$ Mpc$^{-1}$, which is in agreement with the 
low-redshift distance ladder measurements of $H_0^{{\rm DL}} = 73.24 \pm 1.74$ km s$^{-1}$ Mpc$^{-1}$ \cite{2016ApJ...826...56R}.

While these results are encouraging, it needs to be noted that the Simsilun simulation
is not a fully relativistic simulation of our Universe, but it relies on the approximation of the Silent Universe \cite{simsulun.paper}.
Other approaches and approximations to the relativistic numerical cosmology, such as the one based on the weak-field limit do not show the phenomenon of emerging spatial curvature \cite{2017arXiv170609309A} and therefore do not provide the solution of the Hubble constant problem.  
At this stage, the phenomenon of the emerging spatial curvature does seem to be a 
viable and attractive explanation of the  Hubble constant problem.
In fact, one can turn the argument around and argue that the presence of the
tension between low and high-redshift measurements is a moderate (indirect) evidence for 
the phenomenon of emerging spatial curvature. 
From the point of view of astronomical observations, we still do not have a direct measurement of the spatial curvature at low-redshifts. Currently, the low-redshift measurements do not provide any direct measurement of the spatial curvature (available constrains merely result from fitting the FLRW geometry to the data, which is not equivalent to a direct measurment). The situation will change in a few years time with the data from the satellite Euclid \cite{2015PhRvL.115j1301R,2017arXiv170701800B}.

In summary, the results presented in this paper
show that the phenomenon of emerging spatial curvature can provide the 
solution the Hubble constant problem and alleviate the tension between the low and high-redshift  measurements, but it will take a few more years of theoretical and observational work before we will be able to confirm with full certainty that this phenomenon does in fact occur in our Universe.

\section*{Acknowledgements}
This work was supported by the Australian Research Council through the Future Fellowship FT140101270. 
The Millennium simulation databases used in this paper and the web application providing online access to them were constructed as part of the activities of the German Astrophysical Virtual Observatory (GAVO). Computational resources used in this work were provided by the ARC (via FT140101270) and the University of Sydney HPC service (Artemis).

\bibliography{shub_arxiv2} 

\begin{thebibliography}{48}
\expandafter\ifx\csname natexlab\endcsname\relax\def\natexlab#1{#1}\fi
\expandafter\ifx\csname bibnamefont\endcsname\relax
  \def\bibnamefont#1{#1}\fi
\expandafter\ifx\csname bibfnamefont\endcsname\relax
  \def\bibfnamefont#1{#1}\fi
\expandafter\ifx\csname citenamefont\endcsname\relax
  \def\citenamefont#1{#1}\fi
\expandafter\ifx\csname url\endcsname\relax
  \def\url#1{\texttt{#1}}\fi
\expandafter\ifx\csname urlprefix\endcsname\relax\def\urlprefix{URL }\fi
\providecommand{\bibinfo}[2]{#2}
\providecommand{\eprint}[2][]{\url{#2}}

\bibitem[{\citenamefont{{Buchert} et~al.}(2016)\citenamefont{{Buchert},
  {Coley}, {Kleinert}, {Roukema}, and {Wiltshire}}}]{2016IJMPD..2530007B}
\bibinfo{author}{\bibfnamefont{T.}~\bibnamefont{{Buchert}}},
  \bibinfo{author}{\bibfnamefont{A.~A.} \bibnamefont{{Coley}}},
  \bibinfo{author}{\bibfnamefont{H.}~\bibnamefont{{Kleinert}}},
  \bibinfo{author}{\bibfnamefont{B.~F.} \bibnamefont{{Roukema}}},
  \bibnamefont{and} \bibinfo{author}{\bibfnamefont{D.~L.}
  \bibnamefont{{Wiltshire}}}, \bibinfo{journal}{Int. J. Mod. Phys. D}
  \textbf{\bibinfo{volume}{25}}, \bibinfo{eid}{1630007-244}
  (\bibinfo{year}{2016}), \eprint{1512.03313}.

\bibitem[{\citenamefont{{Riess} et~al.}(2016)\citenamefont{{Riess}, {Macri},
  {Hoffmann}, {Scolnic}, {Casertano}, {Filippenko}, {Tucker}, {Reid}, {Jones},
  {Silverman} et~al.}}]{2016ApJ...826...56R}
\bibinfo{author}{\bibfnamefont{A.~G.} \bibnamefont{{Riess}}},
  \bibinfo{author}{\bibfnamefont{L.~M.} \bibnamefont{{Macri}}},
  \bibinfo{author}{\bibfnamefont{S.~L.} \bibnamefont{{Hoffmann}}},
  \bibinfo{author}{\bibfnamefont{D.}~\bibnamefont{{Scolnic}}},
  \bibinfo{author}{\bibfnamefont{S.}~\bibnamefont{{Casertano}}},
  \bibinfo{author}{\bibfnamefont{A.~V.} \bibnamefont{{Filippenko}}},
  \bibinfo{author}{\bibfnamefont{B.~E.} \bibnamefont{{Tucker}}},
  \bibinfo{author}{\bibfnamefont{M.~J.} \bibnamefont{{Reid}}},
  \bibinfo{author}{\bibfnamefont{D.~O.} \bibnamefont{{Jones}}},
  \bibinfo{author}{\bibfnamefont{J.~M.} \bibnamefont{{Silverman}}},
  \bibnamefont{et~al.}, \bibinfo{journal}{Astroph. J.}
  \textbf{\bibinfo{volume}{826}}, \bibinfo{eid}{56} (\bibinfo{year}{2016}),
  \eprint{1604.01424}.

\bibitem[{\citenamefont{{Planck Collaboration}
  et~al.}(2016)\citenamefont{{Planck Collaboration}, {Ade}, {Aghanim},
  {Arnaud}, {Ashdown}, {Aumont}, {Baccigalupi}, {Banday}, {Barreiro},
  {Bartlett} et~al.}}]{2016A&A...594A..13P}
\bibinfo{author}{\bibnamefont{{Planck Collaboration}}},
  \bibinfo{author}{\bibfnamefont{P.~A.~R.} \bibnamefont{{Ade}}},
  \bibinfo{author}{\bibfnamefont{N.}~\bibnamefont{{Aghanim}}},
  \bibinfo{author}{\bibfnamefont{M.}~\bibnamefont{{Arnaud}}},
  \bibinfo{author}{\bibfnamefont{M.}~\bibnamefont{{Ashdown}}},
  \bibinfo{author}{\bibfnamefont{J.}~\bibnamefont{{Aumont}}},
  \bibinfo{author}{\bibfnamefont{C.}~\bibnamefont{{Baccigalupi}}},
  \bibinfo{author}{\bibfnamefont{A.~J.} \bibnamefont{{Banday}}},
  \bibinfo{author}{\bibfnamefont{R.~B.} \bibnamefont{{Barreiro}}},
  \bibinfo{author}{\bibfnamefont{J.~G.} \bibnamefont{{Bartlett}}},
  \bibnamefont{et~al.}, \bibinfo{journal}{Astron. Astrophys.}
  \textbf{\bibinfo{volume}{594}}, \bibinfo{eid}{A13} (\bibinfo{year}{2016}),
  \eprint{1502.01589}.

\bibitem[{\citenamefont{{Guth}}(1981)}]{1981PhRvD..23..347G}
\bibinfo{author}{\bibfnamefont{A.~H.} \bibnamefont{{Guth}}},
  \bibinfo{journal}{Phys. Rev. D} \textbf{\bibinfo{volume}{23}},
  \bibinfo{pages}{347} (\bibinfo{year}{1981}).

\bibitem[{\citenamefont{{Buchert} et~al.}(2015)\citenamefont{{Buchert},
  {Carfora}, {Ellis}, {Kolb}, {MacCallum}, {Ostrowski}, {R{\"a}s{\"a}nen},
  {Roukema}, {Andersson}, {Coley} et~al.}}]{Buchert:2015iva}
\bibinfo{author}{\bibfnamefont{T.}~\bibnamefont{{Buchert}}},
  \bibinfo{author}{\bibfnamefont{M.}~\bibnamefont{{Carfora}}},
  \bibinfo{author}{\bibfnamefont{G.~F.~R.} \bibnamefont{{Ellis}}},
  \bibinfo{author}{\bibfnamefont{E.~W.} \bibnamefont{{Kolb}}},
  \bibinfo{author}{\bibfnamefont{M.~A.~H.} \bibnamefont{{MacCallum}}},
  \bibinfo{author}{\bibfnamefont{J.~J.} \bibnamefont{{Ostrowski}}},
  \bibinfo{author}{\bibfnamefont{S.}~\bibnamefont{{R{\"a}s{\"a}nen}}},
  \bibinfo{author}{\bibfnamefont{B.~F.} \bibnamefont{{Roukema}}},
  \bibinfo{author}{\bibfnamefont{L.}~\bibnamefont{{Andersson}}},
  \bibinfo{author}{\bibfnamefont{A.~A.} \bibnamefont{{Coley}}},
  \bibnamefont{et~al.}, \bibinfo{journal}{Class. Quant. Grav.}
  \textbf{\bibinfo{volume}{32}}, \bibinfo{eid}{215021} (\bibinfo{year}{2015}),
  \eprint{1505.07800}.

\bibitem[{\citenamefont{{Buchert} and {Carfora}}(2008)}]{2008CQGra..25s5001B}
\bibinfo{author}{\bibfnamefont{T.}~\bibnamefont{{Buchert}}} \bibnamefont{and}
  \bibinfo{author}{\bibfnamefont{M.}~\bibnamefont{{Carfora}}},
  \bibinfo{journal}{Classical and Quantum Gravity}
  \textbf{\bibinfo{volume}{25}}, \bibinfo{eid}{195001} (\bibinfo{year}{2008}),
  \eprint{0803.1401}.

\bibitem[{\citenamefont{{Roy} et~al.}(2011)\citenamefont{{Roy}, {Buchert},
  {Carloni}, and {Obadia}}}]{2011CQGra..28p5004R}
\bibinfo{author}{\bibfnamefont{X.}~\bibnamefont{{Roy}}},
  \bibinfo{author}{\bibfnamefont{T.}~\bibnamefont{{Buchert}}},
  \bibinfo{author}{\bibfnamefont{S.}~\bibnamefont{{Carloni}}},
  \bibnamefont{and} \bibinfo{author}{\bibfnamefont{N.}~\bibnamefont{{Obadia}}},
  \bibinfo{journal}{Classical and Quantum Gravity}
  \textbf{\bibinfo{volume}{28}}, \bibinfo{eid}{165004} (\bibinfo{year}{2011}),
  \eprint{1103.1146}.

\bibitem[{\citenamefont{{R{\"a}s{\"a}nen}
  et~al.}(2015)\citenamefont{{R{\"a}s{\"a}nen}, {Bolejko}, and
  {Finoguenov}}}]{2015PhRvL.115j1301R}
\bibinfo{author}{\bibfnamefont{S.}~\bibnamefont{{R{\"a}s{\"a}nen}}},
  \bibinfo{author}{\bibfnamefont{K.}~\bibnamefont{{Bolejko}}},
  \bibnamefont{and}
  \bibinfo{author}{\bibfnamefont{A.}~\bibnamefont{{Finoguenov}}},
  \bibinfo{journal}{Phys. Rev. Lett.} \textbf{\bibinfo{volume}{115}},
  \bibinfo{eid}{101301} (\bibinfo{year}{2015}), \eprint{1412.4976}.

\bibitem[{\citenamefont{{Bolejko} and
  {Korzy{\'n}ski}}(2017)}]{2017IJMPD..2630011B}
\bibinfo{author}{\bibfnamefont{K.}~\bibnamefont{{Bolejko}}} \bibnamefont{and}
  \bibinfo{author}{\bibfnamefont{M.}~\bibnamefont{{Korzy{\'n}ski}}},
  \bibinfo{journal}{International Journal of Modern Physics D}
  \textbf{\bibinfo{volume}{26}}, \bibinfo{eid}{1730011} (\bibinfo{year}{2017}),
  \eprint{1612.08222}.

\bibitem[{\citenamefont{{L{\"o}ffler} et~al.}(2012)\citenamefont{{L{\"o}ffler},
  {Faber}, {Bentivegna}, {Bode}, {Diener}, {Haas}, {Hinder}, {Mundim}, {Ott},
  {Schnetter} et~al.}}]{2012CQGra..29k5001L}
\bibinfo{author}{\bibfnamefont{F.}~\bibnamefont{{L{\"o}ffler}}},
  \bibinfo{author}{\bibfnamefont{J.}~\bibnamefont{{Faber}}},
  \bibinfo{author}{\bibfnamefont{E.}~\bibnamefont{{Bentivegna}}},
  \bibinfo{author}{\bibfnamefont{T.}~\bibnamefont{{Bode}}},
  \bibinfo{author}{\bibfnamefont{P.}~\bibnamefont{{Diener}}},
  \bibinfo{author}{\bibfnamefont{R.}~\bibnamefont{{Haas}}},
  \bibinfo{author}{\bibfnamefont{I.}~\bibnamefont{{Hinder}}},
  \bibinfo{author}{\bibfnamefont{B.~C.} \bibnamefont{{Mundim}}},
  \bibinfo{author}{\bibfnamefont{C.~D.} \bibnamefont{{Ott}}},
  \bibinfo{author}{\bibfnamefont{E.}~\bibnamefont{{Schnetter}}},
  \bibnamefont{et~al.}, \bibinfo{journal}{Classical and Quantum Gravity}
  \textbf{\bibinfo{volume}{29}}, \bibinfo{eid}{115001} (\bibinfo{year}{2012}),
  \eprint{1111.3344}.

\bibitem[{\citenamefont{Bentivegna and Bruni}(2016)}]{Bentivegna:2015flc}
\bibinfo{author}{\bibfnamefont{E.}~\bibnamefont{Bentivegna}} \bibnamefont{and}
  \bibinfo{author}{\bibfnamefont{M.}~\bibnamefont{Bruni}},
  \bibinfo{journal}{Phys. Rev. Lett.} \textbf{\bibinfo{volume}{116}},
  \bibinfo{pages}{251302} (\bibinfo{year}{2016}), \eprint{1511.05124}.

\bibitem[{\citenamefont{Mertens et~al.}(2016)\citenamefont{Mertens, Giblin, and
  Starkman}}]{Mertens:2015ttp}
\bibinfo{author}{\bibfnamefont{J.~B.} \bibnamefont{Mertens}},
  \bibinfo{author}{\bibfnamefont{J.~T.} \bibnamefont{Giblin}},
  \bibnamefont{and} \bibinfo{author}{\bibfnamefont{G.~D.}
  \bibnamefont{Starkman}}, \bibinfo{journal}{Phys. Rev. D}
  \textbf{\bibinfo{volume}{93}}, \bibinfo{pages}{124059}
  (\bibinfo{year}{2016}), \eprint{1511.01106}.

\bibitem[{\citenamefont{{Macpherson} et~al.}(2017)\citenamefont{{Macpherson},
  {Lasky}, and {Price}}}]{2017PhRvD..95f4028M}
\bibinfo{author}{\bibfnamefont{H.~J.} \bibnamefont{{Macpherson}}},
  \bibinfo{author}{\bibfnamefont{P.~D.} \bibnamefont{{Lasky}}},
  \bibnamefont{and} \bibinfo{author}{\bibfnamefont{D.~J.}
  \bibnamefont{{Price}}}, \bibinfo{journal}{Phys. Rev. D}
  \textbf{\bibinfo{volume}{95}}, \bibinfo{eid}{064028} (\bibinfo{year}{2017}),
  \eprint{1611.05447}.

\bibitem[{\citenamefont{{Adamek} et~al.}(2017)\citenamefont{{Adamek},
  {Clarkson}, {Daverio}, {Durrer}, and {Kunz}}}]{2017arXiv170609309A}
\bibinfo{author}{\bibfnamefont{J.}~\bibnamefont{{Adamek}}},
  \bibinfo{author}{\bibfnamefont{C.}~\bibnamefont{{Clarkson}}},
  \bibinfo{author}{\bibfnamefont{D.}~\bibnamefont{{Daverio}}},
  \bibinfo{author}{\bibfnamefont{R.}~\bibnamefont{{Durrer}}}, \bibnamefont{and}
  \bibinfo{author}{\bibfnamefont{M.}~\bibnamefont{{Kunz}}},
  \bibinfo{journal}{ArXiv e-prints}  (\bibinfo{year}{2017}),
  \eprint{1706.09309}.

\bibitem[{\citenamefont{{Bolejko}}(2018{\natexlab{a}})}]{simsulun.paper}
\bibinfo{author}{\bibfnamefont{K.}~\bibnamefont{{Bolejko}}},
  \bibinfo{journal}{Classical and Quantum Gravity}
  \textbf{\bibinfo{volume}{35}}, \bibinfo{eid}{024003}
  (\bibinfo{year}{2018}{\natexlab{a}}), \eprint{1708.09143}.

\bibitem[{\citenamefont{{Hinshaw} et~al.}(2013)\citenamefont{{Hinshaw},
  {Larson}, {Komatsu}, {Spergel}, {Bennett}, {Dunkley}, {Nolta}, {Halpern},
  {Hill}, {Odegard} et~al.}}]{2013ApJS..208...19H}
\bibinfo{author}{\bibfnamefont{G.}~\bibnamefont{{Hinshaw}}},
  \bibinfo{author}{\bibfnamefont{D.}~\bibnamefont{{Larson}}},
  \bibinfo{author}{\bibfnamefont{E.}~\bibnamefont{{Komatsu}}},
  \bibinfo{author}{\bibfnamefont{D.~N.} \bibnamefont{{Spergel}}},
  \bibinfo{author}{\bibfnamefont{C.~L.} \bibnamefont{{Bennett}}},
  \bibinfo{author}{\bibfnamefont{J.}~\bibnamefont{{Dunkley}}},
  \bibinfo{author}{\bibfnamefont{M.~R.} \bibnamefont{{Nolta}}},
  \bibinfo{author}{\bibfnamefont{M.}~\bibnamefont{{Halpern}}},
  \bibinfo{author}{\bibfnamefont{R.~S.} \bibnamefont{{Hill}}},
  \bibinfo{author}{\bibfnamefont{N.}~\bibnamefont{{Odegard}}},
  \bibnamefont{et~al.}, \bibinfo{journal}{Astroph. J. Supp.}
  \textbf{\bibinfo{volume}{208}}, \bibinfo{eid}{19} (\bibinfo{year}{2013}),
  \eprint{1212.5226}.

\bibitem[{\citenamefont{{Ellis}}(1971)}]{1971grc..conf..104E}
\bibinfo{author}{\bibfnamefont{G.~F.~R.} \bibnamefont{{Ellis}}}, in
  \emph{\bibinfo{booktitle}{General Relativity and Cosmology}}, edited by
  \bibinfo{editor}{\bibfnamefont{R.~K.} \bibnamefont{{Sachs}}}
  (\bibinfo{year}{1971}), pp. \bibinfo{pages}{104--182}.

\bibitem[{\citenamefont{{Ellis}}(2009)}]{2009GReGr..41..581E}
\bibinfo{author}{\bibfnamefont{G.~F.~R.} \bibnamefont{{Ellis}}},
  \bibinfo{journal}{General Relativity and Gravitation}
  \textbf{\bibinfo{volume}{41}}, \bibinfo{pages}{581} (\bibinfo{year}{2009}).

\bibitem[{\citenamefont{{Bruni} et~al.}(1995)\citenamefont{{Bruni},
  {Matarrese}, and {Pantano}}}]{1995ApJ...445..958B}
\bibinfo{author}{\bibfnamefont{M.}~\bibnamefont{{Bruni}}},
  \bibinfo{author}{\bibfnamefont{S.}~\bibnamefont{{Matarrese}}},
  \bibnamefont{and}
  \bibinfo{author}{\bibfnamefont{O.}~\bibnamefont{{Pantano}}},
  \bibinfo{journal}{Astroph. J.} \textbf{\bibinfo{volume}{445}},
  \bibinfo{pages}{958} (\bibinfo{year}{1995}), \eprint{astro-ph/9406068}.

\bibitem[{\citenamefont{{van Elst} et~al.}(1997)\citenamefont{{van Elst},
  {Uggla}, {Lesame}, {Ellis}, and {Maartens}}}]{1997CQGra..14.1151V}
\bibinfo{author}{\bibfnamefont{H.}~\bibnamefont{{van Elst}}},
  \bibinfo{author}{\bibfnamefont{C.}~\bibnamefont{{Uggla}}},
  \bibinfo{author}{\bibfnamefont{W.~M.} \bibnamefont{{Lesame}}},
  \bibinfo{author}{\bibfnamefont{G.~F.~R.} \bibnamefont{{Ellis}}},
  \bibnamefont{and}
  \bibinfo{author}{\bibfnamefont{R.}~\bibnamefont{{Maartens}}},
  \bibinfo{journal}{Classical and Quantum Gravity}
  \textbf{\bibinfo{volume}{14}}, \bibinfo{pages}{1151} (\bibinfo{year}{1997}),
  \eprint{gr-qc/9611002}.

\bibitem[{\citenamefont{{Maartens} et~al.}(1997)\citenamefont{{Maartens},
  {Lesame}, and {Ellis}}}]{1997PhRvD..55.5219M}
\bibinfo{author}{\bibfnamefont{R.}~\bibnamefont{{Maartens}}},
  \bibinfo{author}{\bibfnamefont{W.~M.} \bibnamefont{{Lesame}}},
  \bibnamefont{and} \bibinfo{author}{\bibfnamefont{G.~F.~R.}
  \bibnamefont{{Ellis}}}, \bibinfo{journal}{Phys. Rev. D}
  \textbf{\bibinfo{volume}{55}}, \bibinfo{pages}{5219} (\bibinfo{year}{1997}),
  \eprint{gr-qc/9703080}.

\bibitem[{\citenamefont{{Springel} et~al.}(2005)\citenamefont{{Springel},
  {White}, {Jenkins}, {Frenk}, {Yoshida}, {Gao}, {Navarro}, {Thacker},
  {Croton}, {Helly} et~al.}}]{2005Natur.435..629S}
\bibinfo{author}{\bibfnamefont{V.}~\bibnamefont{{Springel}}},
  \bibinfo{author}{\bibfnamefont{S.~D.~M.} \bibnamefont{{White}}},
  \bibinfo{author}{\bibfnamefont{A.}~\bibnamefont{{Jenkins}}},
  \bibinfo{author}{\bibfnamefont{C.~S.} \bibnamefont{{Frenk}}},
  \bibinfo{author}{\bibfnamefont{N.}~\bibnamefont{{Yoshida}}},
  \bibinfo{author}{\bibfnamefont{L.}~\bibnamefont{{Gao}}},
  \bibinfo{author}{\bibfnamefont{J.}~\bibnamefont{{Navarro}}},
  \bibinfo{author}{\bibfnamefont{R.}~\bibnamefont{{Thacker}}},
  \bibinfo{author}{\bibfnamefont{D.}~\bibnamefont{{Croton}}},
  \bibinfo{author}{\bibfnamefont{J.}~\bibnamefont{{Helly}}},
  \bibnamefont{et~al.}, \bibinfo{journal}{Nature}
  \textbf{\bibinfo{volume}{435}}, \bibinfo{pages}{629} (\bibinfo{year}{2005}),
  \eprint{astro-ph/0504097}.

\bibitem[{\citenamefont{{Boylan-Kolchin}
  et~al.}(2009)\citenamefont{{Boylan-Kolchin}, {Springel}, {White}, {Jenkins},
  and {Lemson}}}]{2009MNRAS.398.1150B}
\bibinfo{author}{\bibfnamefont{M.}~\bibnamefont{{Boylan-Kolchin}}},
  \bibinfo{author}{\bibfnamefont{V.}~\bibnamefont{{Springel}}},
  \bibinfo{author}{\bibfnamefont{S.~D.~M.} \bibnamefont{{White}}},
  \bibinfo{author}{\bibfnamefont{A.}~\bibnamefont{{Jenkins}}},
  \bibnamefont{and} \bibinfo{author}{\bibfnamefont{G.}~\bibnamefont{{Lemson}}},
  \bibinfo{journal}{Mon. Not. R. Astron. Soc.} \textbf{\bibinfo{volume}{398}},
  \bibinfo{pages}{1150} (\bibinfo{year}{2009}), \eprint{0903.3041}.

\bibitem[{\citenamefont{{Guo} et~al.}(2013)\citenamefont{{Guo}, {White},
  {Angulo}, {Henriques}, {Lemson}, {Boylan-Kolchin}, {Thomas}, and
  {Short}}}]{2013MNRAS.428.1351G}
\bibinfo{author}{\bibfnamefont{Q.}~\bibnamefont{{Guo}}},
  \bibinfo{author}{\bibfnamefont{S.}~\bibnamefont{{White}}},
  \bibinfo{author}{\bibfnamefont{R.~E.} \bibnamefont{{Angulo}}},
  \bibinfo{author}{\bibfnamefont{B.}~\bibnamefont{{Henriques}}},
  \bibinfo{author}{\bibfnamefont{G.}~\bibnamefont{{Lemson}}},
  \bibinfo{author}{\bibfnamefont{M.}~\bibnamefont{{Boylan-Kolchin}}},
  \bibinfo{author}{\bibfnamefont{P.}~\bibnamefont{{Thomas}}}, \bibnamefont{and}
  \bibinfo{author}{\bibfnamefont{C.}~\bibnamefont{{Short}}},
  \bibinfo{journal}{Mon. Not. R. Astron. Soc.} \textbf{\bibinfo{volume}{428}},
  \bibinfo{pages}{1351} (\bibinfo{year}{2013}), \eprint{1206.0052}.

\bibitem[{\citenamefont{{Bolejko}}(2018{\natexlab{b}})}]{2017arXiv171204041B}
\bibinfo{author}{\bibfnamefont{K.}~\bibnamefont{{Bolejko}}},
  \bibinfo{journal}{Phys. Rev. D} \textbf{\bibinfo{volume}{97}}
  (\bibinfo{year}{2018}{\natexlab{b}}).

\bibitem[{\citenamefont{{Krasi{\'n}ski}
  et~al.}(2010)\citenamefont{{Krasi{\'n}ski}, {Hellaby}, {Bolejko}, and
  {C{\'e}l{\'e}rier}}}]{2010GReGr..42.2453K}
\bibinfo{author}{\bibfnamefont{A.}~\bibnamefont{{Krasi{\'n}ski}}},
  \bibinfo{author}{\bibfnamefont{C.}~\bibnamefont{{Hellaby}}},
  \bibinfo{author}{\bibfnamefont{K.}~\bibnamefont{{Bolejko}}},
  \bibnamefont{and} \bibinfo{author}{\bibfnamefont{M.-N.}
  \bibnamefont{{C{\'e}l{\'e}rier}}}, \bibinfo{journal}{Gen. Rel. Grav.}
  \textbf{\bibinfo{volume}{42}}, \bibinfo{pages}{2453} (\bibinfo{year}{2010}),
  \eprint{0903.4070}.

\bibitem[{\citenamefont{{Sachs}}(1961)}]{1961RSPSA.264..309S}
\bibinfo{author}{\bibfnamefont{R.}~\bibnamefont{{Sachs}}},
  \bibinfo{journal}{Proceedings of the Royal Society of London Series A}
  \textbf{\bibinfo{volume}{264}}, \bibinfo{pages}{309} (\bibinfo{year}{1961}).

\bibitem[{\citenamefont{{Bolejko} and {Ferreira}}(2012)}]{2012JCAP...05..003B}
\bibinfo{author}{\bibfnamefont{K.}~\bibnamefont{{Bolejko}}} \bibnamefont{and}
  \bibinfo{author}{\bibfnamefont{P.~G.} \bibnamefont{{Ferreira}}},
  \bibinfo{journal}{J. Cosmol. Astropart. Phys.} \textbf{\bibinfo{volume}{5}},
  \bibinfo{eid}{003} (\bibinfo{year}{2012}), \eprint{1204.0909}.

\bibitem[{\citenamefont{{R{\"a}s{\"a}nen}}(2010)}]{2010JCAP...03..018R}
\bibinfo{author}{\bibfnamefont{S.}~\bibnamefont{{R{\"a}s{\"a}nen}}},
  \bibinfo{journal}{J. Cosmol. Astropart. Phys.} \textbf{\bibinfo{volume}{3}},
  \bibinfo{eid}{018} (\bibinfo{year}{2010}), \eprint{0912.3370}.

\bibitem[{\citenamefont{{Suzuki} et~al.}(2012)\citenamefont{{Suzuki}, {Rubin},
  {Lidman}, {Aldering}, {Amanullah}, {Barbary}, {Barrientos}, {Botyanszki},
  {Brodwin}, {Connolly} et~al.}}]{2012ApJ...746...85S}
\bibinfo{author}{\bibfnamefont{N.}~\bibnamefont{{Suzuki}}},
  \bibinfo{author}{\bibfnamefont{D.}~\bibnamefont{{Rubin}}},
  \bibinfo{author}{\bibfnamefont{C.}~\bibnamefont{{Lidman}}},
  \bibinfo{author}{\bibfnamefont{G.}~\bibnamefont{{Aldering}}},
  \bibinfo{author}{\bibfnamefont{R.}~\bibnamefont{{Amanullah}}},
  \bibinfo{author}{\bibfnamefont{K.}~\bibnamefont{{Barbary}}},
  \bibinfo{author}{\bibfnamefont{L.~F.} \bibnamefont{{Barrientos}}},
  \bibinfo{author}{\bibfnamefont{J.}~\bibnamefont{{Botyanszki}}},
  \bibinfo{author}{\bibfnamefont{M.}~\bibnamefont{{Brodwin}}},
  \bibinfo{author}{\bibfnamefont{N.}~\bibnamefont{{Connolly}}},
  \bibnamefont{et~al.}, \bibinfo{journal}{Astroph. J.}
  \textbf{\bibinfo{volume}{746}}, \bibinfo{eid}{85} (\bibinfo{year}{2012}),
  \eprint{1105.3470}.

\bibitem[{\citenamefont{{Dyer} and {Roeder}}(1972)}]{1972ApJ...174L.115D}
\bibinfo{author}{\bibfnamefont{C.~C.} \bibnamefont{{Dyer}}} \bibnamefont{and}
  \bibinfo{author}{\bibfnamefont{R.~C.} \bibnamefont{{Roeder}}},
  \bibinfo{journal}{Astroph. J.} \textbf{\bibinfo{volume}{174}},
  \bibinfo{pages}{L115} (\bibinfo{year}{1972}).

\bibitem[{\citenamefont{{Dyer} and {Roeder}}(1973)}]{1973ApJ...180L..31D}
\bibinfo{author}{\bibfnamefont{C.~C.} \bibnamefont{{Dyer}}} \bibnamefont{and}
  \bibinfo{author}{\bibfnamefont{R.~C.} \bibnamefont{{Roeder}}},
  \bibinfo{journal}{Astroph. J.} \textbf{\bibinfo{volume}{180}},
  \bibinfo{pages}{L31} (\bibinfo{year}{1973}).

\bibitem[{\citenamefont{{Bolejko}}(2011)}]{2011MNRAS.412.1937B}
\bibinfo{author}{\bibfnamefont{K.}~\bibnamefont{{Bolejko}}},
  \bibinfo{journal}{Mon. Not. R. Astron. Soc.} \textbf{\bibinfo{volume}{412}},
  \bibinfo{pages}{1937} (\bibinfo{year}{2011}), \eprint{1011.3876}.

\bibitem[{\citenamefont{{Clarkson} et~al.}(2012)\citenamefont{{Clarkson},
  {Ellis}, {Faltenbacher}, {Maartens}, {Umeh}, and
  {Uzan}}}]{2012MNRAS.426.1121C}
\bibinfo{author}{\bibfnamefont{C.}~\bibnamefont{{Clarkson}}},
  \bibinfo{author}{\bibfnamefont{G.~F.~R.} \bibnamefont{{Ellis}}},
  \bibinfo{author}{\bibfnamefont{A.}~\bibnamefont{{Faltenbacher}}},
  \bibinfo{author}{\bibfnamefont{R.}~\bibnamefont{{Maartens}}},
  \bibinfo{author}{\bibfnamefont{O.}~\bibnamefont{{Umeh}}}, \bibnamefont{and}
  \bibinfo{author}{\bibfnamefont{J.-P.} \bibnamefont{{Uzan}}},
  \bibinfo{journal}{Mon. Not. R. Astron. Soc.} \textbf{\bibinfo{volume}{426}},
  \bibinfo{pages}{1121} (\bibinfo{year}{2012}), \eprint{1109.2484}.

\bibitem[{\citenamefont{{Meures} and {Bruni}}(2012)}]{2012MNRAS.419.1937M}
\bibinfo{author}{\bibfnamefont{N.}~\bibnamefont{{Meures}}} \bibnamefont{and}
  \bibinfo{author}{\bibfnamefont{M.}~\bibnamefont{{Bruni}}},
  \bibinfo{journal}{Mon. Not. R. Astron. Soc.} \textbf{\bibinfo{volume}{419}},
  \bibinfo{pages}{1937} (\bibinfo{year}{2012}), \eprint{1107.4433}.

\bibitem[{\citenamefont{{Vanderveld} et~al.}(2008)\citenamefont{{Vanderveld},
  {Flanagan}, and {Wasserman}}}]{2008PhRvD..78h3511V}
\bibinfo{author}{\bibfnamefont{R.~A.} \bibnamefont{{Vanderveld}}},
  \bibinfo{author}{\bibfnamefont{{\'E}.~{\'E}.} \bibnamefont{{Flanagan}}},
  \bibnamefont{and}
  \bibinfo{author}{\bibfnamefont{I.}~\bibnamefont{{Wasserman}}},
  \bibinfo{journal}{Phys. Rev. D} \textbf{\bibinfo{volume}{78}},
  \bibinfo{eid}{083511} (\bibinfo{year}{2008}), \eprint{0808.1080}.

\bibitem[{\citenamefont{{Kainulainen} and {Marra}}(2009)}]{2009PhRvD..80l3020K}
\bibinfo{author}{\bibfnamefont{K.}~\bibnamefont{{Kainulainen}}}
  \bibnamefont{and} \bibinfo{author}{\bibfnamefont{V.}~\bibnamefont{{Marra}}},
  \bibinfo{journal}{Phys. Rev. D} \textbf{\bibinfo{volume}{80}},
  \bibinfo{eid}{123020} (\bibinfo{year}{2009}), \eprint{0909.0822}.

\bibitem[{\citenamefont{{R{\"a}s{\"a}nen}}(2009)}]{2009JCAP...02..011R}
\bibinfo{author}{\bibfnamefont{S.}~\bibnamefont{{R{\"a}s{\"a}nen}}},
  \bibinfo{journal}{J. Cosmol. Astropart. Phys.} \textbf{\bibinfo{volume}{2}},
  \bibinfo{eid}{011} (\bibinfo{year}{2009}), \eprint{0812.2872}.

\bibitem[{\citenamefont{{Kaiser} and {Peacock}}(2016)}]{2016MNRAS.455.4518K}
\bibinfo{author}{\bibfnamefont{N.}~\bibnamefont{{Kaiser}}} \bibnamefont{and}
  \bibinfo{author}{\bibfnamefont{J.~A.} \bibnamefont{{Peacock}}},
  \bibinfo{journal}{Mon. Not. R. Astron. Soc.} \textbf{\bibinfo{volume}{455}},
  \bibinfo{pages}{4518} (\bibinfo{year}{2016}), \eprint{1503.08506}.

\bibitem[{\citenamefont{{Tammann}}(2006)}]{2006RvMA...19....1T}
\bibinfo{author}{\bibfnamefont{G.~A.} \bibnamefont{{Tammann}}}, in
  \emph{\bibinfo{booktitle}{Reviews in Modern Astronomy}}, edited by
  \bibinfo{editor}{\bibfnamefont{S.}~\bibnamefont{{Roeser}}}
  (\bibinfo{year}{2006}), vol.~\bibinfo{volume}{19} of
  \emph{\bibinfo{series}{Reviews in Modern Astronomy}}, p.~\bibinfo{pages}{1},
  \eprint{astro-ph/0512584}.

\bibitem[{\citenamefont{{Freedman} et~al.}(2001)\citenamefont{{Freedman},
  {Madore}, {Gibson}, {Ferrarese}, {Kelson}, {Sakai}, {Mould}, {Kennicutt},
  {Ford}, {Graham} et~al.}}]{2001ApJ...553...47F}
\bibinfo{author}{\bibfnamefont{W.~L.} \bibnamefont{{Freedman}}},
  \bibinfo{author}{\bibfnamefont{B.~F.} \bibnamefont{{Madore}}},
  \bibinfo{author}{\bibfnamefont{B.~K.} \bibnamefont{{Gibson}}},
  \bibinfo{author}{\bibfnamefont{L.}~\bibnamefont{{Ferrarese}}},
  \bibinfo{author}{\bibfnamefont{D.~D.} \bibnamefont{{Kelson}}},
  \bibinfo{author}{\bibfnamefont{S.}~\bibnamefont{{Sakai}}},
  \bibinfo{author}{\bibfnamefont{J.~R.} \bibnamefont{{Mould}}},
  \bibinfo{author}{\bibfnamefont{R.~C.} \bibnamefont{{Kennicutt}},
  \bibfnamefont{Jr.}}, \bibinfo{author}{\bibfnamefont{H.~C.}
  \bibnamefont{{Ford}}}, \bibinfo{author}{\bibfnamefont{J.~A.}
  \bibnamefont{{Graham}}}, \bibnamefont{et~al.}, \bibinfo{journal}{Astroph. J.}
  \textbf{\bibinfo{volume}{553}}, \bibinfo{pages}{47} (\bibinfo{year}{2001}),
  \eprint{astro-ph/0012376}.

\bibitem[{\citenamefont{{Komatsu} et~al.}(2011)\citenamefont{{Komatsu},
  {Smith}, {Dunkley}, {Bennett}, {Gold}, {Hinshaw}, {Jarosik}, {Larson},
  {Nolta}, {Page} et~al.}}]{2011ApJS..192...18K}
\bibinfo{author}{\bibfnamefont{E.}~\bibnamefont{{Komatsu}}},
  \bibinfo{author}{\bibfnamefont{K.~M.} \bibnamefont{{Smith}}},
  \bibinfo{author}{\bibfnamefont{J.}~\bibnamefont{{Dunkley}}},
  \bibinfo{author}{\bibfnamefont{C.~L.} \bibnamefont{{Bennett}}},
  \bibinfo{author}{\bibfnamefont{B.}~\bibnamefont{{Gold}}},
  \bibinfo{author}{\bibfnamefont{G.}~\bibnamefont{{Hinshaw}}},
  \bibinfo{author}{\bibfnamefont{N.}~\bibnamefont{{Jarosik}}},
  \bibinfo{author}{\bibfnamefont{D.}~\bibnamefont{{Larson}}},
  \bibinfo{author}{\bibfnamefont{M.~R.} \bibnamefont{{Nolta}}},
  \bibinfo{author}{\bibfnamefont{L.}~\bibnamefont{{Page}}},
  \bibnamefont{et~al.}, \bibinfo{journal}{Astroph. J. Supp.}
  \textbf{\bibinfo{volume}{192}}, \bibinfo{eid}{18} (\bibinfo{year}{2011}),
  \eprint{1001.4538}.

\bibitem[{\citenamefont{{Riess} et~al.}(2011)\citenamefont{{Riess}, {Macri},
  {Casertano}, {Lampeitl}, {Ferguson}, {Filippenko}, {Jha}, {Li}, and
  {Chornock}}}]{2011ApJ...730..119R}
\bibinfo{author}{\bibfnamefont{A.~G.} \bibnamefont{{Riess}}},
  \bibinfo{author}{\bibfnamefont{L.}~\bibnamefont{{Macri}}},
  \bibinfo{author}{\bibfnamefont{S.}~\bibnamefont{{Casertano}}},
  \bibinfo{author}{\bibfnamefont{H.}~\bibnamefont{{Lampeitl}}},
  \bibinfo{author}{\bibfnamefont{H.~C.} \bibnamefont{{Ferguson}}},
  \bibinfo{author}{\bibfnamefont{A.~V.} \bibnamefont{{Filippenko}}},
  \bibinfo{author}{\bibfnamefont{S.~W.} \bibnamefont{{Jha}}},
  \bibinfo{author}{\bibfnamefont{W.}~\bibnamefont{{Li}}}, \bibnamefont{and}
  \bibinfo{author}{\bibfnamefont{R.}~\bibnamefont{{Chornock}}},
  \bibinfo{journal}{Astroph. J.} \textbf{\bibinfo{volume}{730}},
  \bibinfo{eid}{119} (\bibinfo{year}{2011}), \eprint{1103.2976}.

\bibitem[{\citenamefont{{Planck Collaboration}
  et~al.}(2014)\citenamefont{{Planck Collaboration}, {Ade}, {Aghanim},
  {Armitage-Caplan}, {Arnaud}, {Ashdown}, {Atrio-Barandela}, {Aumont},
  {Baccigalupi}, {Banday} et~al.}}]{2014A&A...571A..16P}
\bibinfo{author}{\bibnamefont{{Planck Collaboration}}},
  \bibinfo{author}{\bibfnamefont{P.~A.~R.} \bibnamefont{{Ade}}},
  \bibinfo{author}{\bibfnamefont{N.}~\bibnamefont{{Aghanim}}},
  \bibinfo{author}{\bibfnamefont{C.}~\bibnamefont{{Armitage-Caplan}}},
  \bibinfo{author}{\bibfnamefont{M.}~\bibnamefont{{Arnaud}}},
  \bibinfo{author}{\bibfnamefont{M.}~\bibnamefont{{Ashdown}}},
  \bibinfo{author}{\bibfnamefont{F.}~\bibnamefont{{Atrio-Barandela}}},
  \bibinfo{author}{\bibfnamefont{J.}~\bibnamefont{{Aumont}}},
  \bibinfo{author}{\bibfnamefont{C.}~\bibnamefont{{Baccigalupi}}},
  \bibinfo{author}{\bibfnamefont{A.~J.} \bibnamefont{{Banday}}},
  \bibnamefont{et~al.}, \bibinfo{journal}{Astron. Astrophys.}
  \textbf{\bibinfo{volume}{571}}, \bibinfo{eid}{A16} (\bibinfo{year}{2014}),
  \eprint{1303.5076}.

\bibitem[{\citenamefont{{Lin} and {Ishak}}(2017)}]{2017PhRvD..96h3532L}
\bibinfo{author}{\bibfnamefont{W.}~\bibnamefont{{Lin}}} \bibnamefont{and}
  \bibinfo{author}{\bibfnamefont{M.}~\bibnamefont{{Ishak}}},
  \bibinfo{journal}{\prd} \textbf{\bibinfo{volume}{96}}, \bibinfo{eid}{083532}
  (\bibinfo{year}{2017}), \eprint{1708.09813}.

\bibitem[{\citenamefont{{Efstathiou}}(2014)}]{2014MNRAS.440.1138E}
\bibinfo{author}{\bibfnamefont{G.}~\bibnamefont{{Efstathiou}}},
  \bibinfo{journal}{Mon. Not. R. Astron. Soc.} \textbf{\bibinfo{volume}{440}},
  \bibinfo{pages}{1138} (\bibinfo{year}{2014}), \eprint{1311.3461}.

\bibitem[{\citenamefont{{Bennett} et~al.}(2014)\citenamefont{{Bennett},
  {Larson}, {Weiland}, and {Hinshaw}}}]{2014ApJ...794..135B}
\bibinfo{author}{\bibfnamefont{C.~L.} \bibnamefont{{Bennett}}},
  \bibinfo{author}{\bibfnamefont{D.}~\bibnamefont{{Larson}}},
  \bibinfo{author}{\bibfnamefont{J.~L.} \bibnamefont{{Weiland}}},
  \bibnamefont{and}
  \bibinfo{author}{\bibfnamefont{G.}~\bibnamefont{{Hinshaw}}},
  \bibinfo{journal}{Astroph. J.} \textbf{\bibinfo{volume}{794}},
  \bibinfo{eid}{135} (\bibinfo{year}{2014}), \eprint{1406.1718}.

\bibitem[{\citenamefont{{Bolejko}}(2017)}]{2017arXiv170701800B}
\bibinfo{author}{\bibfnamefont{K.}~\bibnamefont{{Bolejko}}},
  \bibinfo{journal}{ArXiv e-prints}  (\bibinfo{year}{2017}),
  \eprint{1707.01800}.

\end{thebibliography}
\bibliographystyle{apsrev}

\end{document}